\documentclass[12pt]{article}
\usepackage{amsfonts,amssymb,amsmath,amscd}
\usepackage{epsf}

\def\d{\partial}

\def\d{\partial}

\def\({\left(}
\def\){\right)}
\def\[{\left[}
\def\]{\right]}

\def\({\left(}
\def\){\right)}

\def\d{\partial}

\def\beq{\begin{equation}}
\def\eeq{\end{equation}}
\def\bea{\begin{eqnarray}}
\def\eea{\end{eqnarray}}
\def\bq{\begin{quote}}
\def\eq{\end{quote}}

\def\d{\partial}

\def\({\left(}
\def\){\right)}

\def\g5{\gamma_5}
\def\gappeq{\mathrel{\rlap {\raise.5ex\hbox{$>$}}
{\lower.5ex\hbox{$\sim$}}}}

\def\lappeq{\mathrel{\rlap{\raise.5ex\hbox{$<$}}
{\lower.5ex\hbox{$\sim$}}}}

\def\Toprel#1\over#2{\mathrel{\mathop{#2}\limits^{#1}}}

\begin{document}

\pagestyle{empty}
\begin{flushright}
{CPHT-RR 058.0904}\\ 
{\ttfamily hep-th/0409305}\\ 
\end{flushright}
\vspace*{5mm}
\begin{center}
{ \Huge Deformation of $p$-adic String Amplitudes 
 \\
 \vspace*{3mm} \Huge in a Magnetic Field}\\
\vspace*{19mm}
{\large Pascal Grange} \\
\vspace{0.5cm}

 {\it Centre de physique th{\'e}orique de l'{\'E}cole polytechnique,\\
 \vspace*{0.2cm}route de Saclay, 91128 Palaiseau Cedex, France}\\
\vspace{0.5cm} 
{\tt{ pascal.grange@cpht.polytechnique.fr}}\\
\vspace*{3cm}

\vspace*{1.5cm}  
{\bf ABSTRACT} \\ \end{center}
\vspace*{5mm}
\noindent

A new term in the $p$-adic world-sheet action is proposed, which couples a 
   constant $B$-field  to the boundary of the world-sheet at
   disk level. The induced deformation of tachyon scattering amplitudes by star-products
 is  derived. This is in agreement with the deformation of effective
 action postulated in recent investigations of noncommutative solitons
   in $p$-adic string theory.

\vspace*{0.5cm}
\noindent

\begin{flushleft} 
December 2004 
\end{flushleft}

\vfill\eject
%\pagestyle{empty}
%\clearpage\mbox{}\clearpage

\setcounter{page}{1}
\pagestyle{plain}

\section{Introduction}

It has been appreciated that allowing
non-Archimedean number fields (for a review, see~\cite{BF}) leads to
modifications of string
theory, thereby yielding exactly solvable models. Given a prime number $p$, one may consider the field
${\rm\bf{Q}}_p$ of $p$-adic numbers, and modify the theory by making the coordinates
$p$-adic\footnote{The $p$-adic norm of a rational number $x$ is
  defined as $|x|_p=p^{-n}$, where $p^n$ is the largest power of $p$ dividing $x$. The
  field ${\rm\bf{Q}}_p$ is defined as the Cauchy completion of
  ${\rm\bf{Q}}$ with the $p$-adic norm. Measure, integration, changes
  of variables and Fourier transform can be defined on ${\rm\bf{Q}}_p$,
whereas derivatives do not find a straightforward analog. This lack of
analogy makes the difference between the exhibition of a
world-sheet Lagrangian 
and a translation work.} on the boundary of the world-sheet.
Other quantities such as actions, amplitudes, target-space coordinates and 
coordinates in the interior of the world-sheet remain usual complex
numbers.  
 Moreover, as shown by Zabrodin~\cite{Zabrodin}, realization of the
 boundary as the projective set of leaves of an
 infinite tree allows for a new picture of $p$-adic string theory at
 disk level. The action of the bosonic string must be rewritten using
 some discrete scheme, which couples the space-time metric to the
 tree:
$$\int_\Sigma d^2\sigma \,\d_\alpha X^\mu \d^\alpha X_\mu \longrightarrow
\sum_{i}\sum_{j(i)}{\Big{(}} X^\mu(j(i))- X^\mu(i)\Big{)}\Big{(} X_\mu(j(i))-
X_\mu(i){\Big{)}}.$$
 Degrees of freedom corresponding to the interior of the world-sheet
  can furthermore be integrated out, leading to a non-local action
  on the $p$-adic boundary~\cite{Zhang,Spokoiny,Parisi}. Scattering
  amplitudes of tachyons can be recovered
  either by computing path integrals with Neumann
  boundary conditions in the discretized scheme, or by saddle-point
  method with the non-local action. The output of both methods is
  the tachyon Lagrangian~\cite{Frampton1,Frampton2,Frampton3,FO,BFW}
$$\mathcal{L}=-\frac{1}{2}\phi p^{-1/2\triangle}\phi+\frac{1}{p+1}\phi ^{p+1},$$
 whose exactness gives appeal to this whole
  approach by non-Archimedean number fields. It provides a toy model
  for string theory, allowing to explicitly
  check the scenario of tachyon condensation~\cite{GS}.\\

 The recent work by Ghoshal~\cite{Ghoshal} investigated noncommutative
 $p$-adic solitons, postulating a deformation of the effective action
 by star-products coming from the $B$-field:
$$\mathcal{L}= -\frac{1}{2}\phi \ast p^{-1/2\triangle}\phi+\frac{1}{p+1}(\ast\phi) ^{p+1}.$$
 This natural
 guess begs for a world-sheet action. First of all, one may ask
 how the $B$-field  couples to the $p$-adic boundary of the disk, and
 subsequently compute the modifications induced by this coupling in
 the scattering amplitudes. To this end, I shall consider the
 $B$-field 
as a magnetic field coupled to the boundary of the disk. This
 will allow for quicker comparison to the previously known non-local
 term involving only the metric. Since I only consider a constant
 $B$-field, inspiration from usual string theory indeed allows for an
 investigation of mere boundary couplings, through
$$\int_\Sigma B =\int_{\d\Sigma} B_{\mu\nu} X^\mu\frac{dX^\nu}{dt}dt.$$ 
 In this letter, I shall propose a natural $p$-adic analog of this
 coupling. Using special functions  on extensions of  ${\rm\bf{Q}}_p$, I shall
   incorporate this additional term in the path integral, and read off
   star-products on scattering amplitudes.

\section{Coupling the $B$-field to the boundary}

When no $B$-field is turned on, the following non-local action on the
$p$-adic boundary of the world-sheet was shown by Zhang~\cite{Zhang} to
lead to the scattering amplitudes previously derived by the means of
discretized Laplacian operator on the whole world-sheet:
$$S[X]= \int_{{\rm\bf Q}_p}du\,\Big{(}g_{\mu\nu}X^\mu(-u)|u|_p
X^\nu(u)\Big{)},$$
$$A_N(k_1,\dots,k_N)=\int DX \exp\(-S[X]+i\sum_{i=1}^N
k_{i,\mu}X^\mu(u_i)\)= \prod_{1\leq i < j\leq N} |u_i-u_j|_p^{ k_{i,\mu}
  g^{\mu\nu} k_{j,\nu}}.$$ 

On dimensional grounds, we expect the same structure as in the
symmetric case, as far as powers of $u$ are concerned, because the
same number of derivatives are present in usual string 
theory. The essential difference comes from the index structure. We
need an antisymmetric kernel in order to write down a non-zero coupling
to $B$. Fortunately, a $p$-adic notion of sign function does exist, though
it is not unique~\cite{Serre,Gelfand} and depends on the choice of a
quadratic extension of the field of $p$-adic numbers.
 Consider the $\epsilon_\tau$-function, defined on the quadratic extension of ${\bf{Q}}_p$
by a non-square $\tau$ as follows: $\epsilon_\tau(x)=1$ if $x$ is the
product of two conjugate numbers in ${\bf{Q}}_p(\sqrt{\tau})$, and
$\epsilon_\tau(x)=-1$ otherwise. It is a multiplicative character, and
for it to contribute a minus sign, we need the property
 $$\epsilon_\tau(x-y)=-\epsilon_\tau(y-x),$$
which reduces to 
$$\epsilon_{\tau}(-1)=-1.$$
This property is however {\it{not}} verified if we take $\tau$ to be a
$(p-1)$-th root of unity, but can be achieved by choosing $\tau$ to be
$p$ or a multiple of it by a $(p-1)$-th root of unity. From now on I
shall just write $\epsilon$ for this sign function, and $|~|$ for the
$p$-adic norm.
 I therefore write the following non-local action on the $p$-adic
boundary in flat space-time, coupled to a constant $B$-field:
$$S[X]=\int_{{\rm\bf Q}_p}du\,\Big{(}\delta_{\mu\nu}X^\mu(-u)|u|
X^\nu(u)+B_{\mu\nu}X^\mu(-u)|u|\epsilon(u) X^\nu(u)\Big{)}.$$

We are now instructed to use this action to compute saddle-point
approximation to the path integral corresponding to the 
scattering of $N$ tachyons. To this end we need the Green function of
the integration kernel we have written down.
 It  is computed by splitting the action between symmetric and
antisymmetric parts, as far as tensors are concerned, because the flat
metric tensor commutes with all the tensors at
hand and their inverses. We denote by $G$ and $\Theta$ the symmetric
and antisymmetric parts of the inverse of $\delta+B$:  
  $$G^{\mu\nu}:=\(\frac{1}{\delta+B}\)^{\mu\alpha}g_{\alpha\beta}\(\frac{1}{\delta-B}\)^{\beta\nu},$$
 $$ \Theta
 ^{\mu\nu}:=-\(\frac{1}{\delta+B}\)^{\mu\alpha}B_{\alpha\beta}
 \(\frac{1}{\delta-B}\)^{\beta\nu}.$$
The Green function is obtained by Fourier transform, taking the same
regularization scheme by a small parameter $s$ as in~\cite{Zhang}:
$$\Delta_{reg} ^{\mu\nu}(x-y)=\int_{{\rm\bf Q}_p} du \,e ^{i(x-y)u}\(G
^{\mu\nu}\frac{1}{|u|^{1+s}}+\Theta^{\mu\nu}\frac{\epsilon(u)}{|u|^{1+s}}\).$$

The problem of the singularity at $s=0$ was dealt with in~\cite{Zhang}
by extracting the finite part in an expansion in powers of $s$. 
This problem does not show up in
the second term we have just induced, and we
 do not need to remove any singularity from it at $s=0$. The sign
 function we use
 can be integrated against functions over $p$-adic numbers, just as in
 Fourier transform. This procedure provides a generalization of
 Gamma-functions:
$$\hat{\Gamma}_p(s):=\int_{{\rm{\bf Q}}_p}du\,  e ^{i2\pi[u]}\epsilon(u)|u|^{s-1}.$$       
 These generalized functions nicely parallel the role played by
 Gamma-functions in~\cite{Zhang}. 
Changing integration variables to $v=(x-y)u$, we obtain
 $$\Delta_{reg} ^{\mu\nu}(x-y)=|x-y|^{-s}\int_{{\rm\bf Q}_p} dv \,e ^{iv}\(G
^{\mu\nu}\frac{1}{|v|^{1+s}}+\Theta^{\mu\nu}\epsilon(x-y)\frac{\epsilon(v)}{|v|^{1+s}}\),$$
which reads in terms of $p$-adic Gamma-functions and generalizations
thereof:
$$\Delta_{reg} ^{\mu\nu}(x-y)= |x-y|^{-s}\Big{(}G^{\mu\nu}\Gamma_p(s) +
\Theta^{\mu\nu}\epsilon(x-y)\hat{\Gamma}_p(s)\Big{)}.$$
On extensions of ${{\rm\bf Q}_p}$ where the sign of $-1$ can be negative, generalized
 Gamma-functions obey the following up to a sign:
$$\hat{\Gamma}_p(s) =\sqrt{\epsilon(-1)} p^{s-1/2},$$
so that, with $\epsilon(-1)=-1$, the regular part of the propagator reads
$$\Delta^{\mu\nu}(x-y)=G^{\mu\nu}\log |x-y| +\frac{i}{\sqrt{p}}\Theta^{\mu\nu}\epsilon(x-y).$$
This completes the derivation  of the Green function needed to
evaluate scattering amplitudes. The guess of the coupling resides
essentially in symmetry properties of the kernel. The choice of a quadratic 
extension of ${\rm{\bf Q}}_p$ with negative sign for -1 is crucial to
obtain a non-zero coupling.

\section{Modification of the scattering amplitudes}
We are now able to compute scattering amplitudes at disk level,
thereby observing the modifications induced by the
$B$-field. Without the $B$-field, the non-local action leads to
amplitudes that are formally the same as in ordinary string theory,
except for integration on ${\rm{\bf Q}}_p$ and prefactors depending on $p$.  Checking whether our action leads to the deformation postulated
in~\cite{Ghoshal} will amount to checking whether the modification of
the scattering amplitudes by the antisymmetric
piece of the propagator neatly arranges into phases that can be
recognized as Fourier transforms of star-products. This is the lesson
of Feynman rules of noncommutative field theory~\cite{SW,GMS}.  

Let us write the  saddle-point
approximation to the following path integral corresponding to
scattering of $N$ tachyons:
$$A_N(k_1,\dots,k_N)=\int DX \exp\(-S[X]+i\sum_{i=1}^N
k_{i,\mu}X^\mu(u_i)\).$$
 Contractions between pairs of scalars using the
propagator derived above yield phases reminiscent of the Moyal vertex in ordinary
string theory~\cite{SW}:
$$A_N(k_1,\dots,k_N)=\prod_{1\leq i < j\leq N}  e ^{\frac{i}{2\sqrt{p}} k_{i,\mu}
  \Theta^{\mu\nu} k_{j,\nu}\epsilon(u_i-u_j)} |u_i-u_j|^{ k_{i,\mu}
  G^{\mu\nu} k_{j,\nu}}.$$
The influence of the $B$-field is completely encoded into phases,
which read as Fourier transform of star-products. We have therefore
established the relationship:\\
$$ \int DX \exp\(-S_{G,\Theta}[X]+ i\sum_{i=1}^N
k_{i,\mu}X^\mu(u_i)\)=$$
$$\prod_{1\leq i < j\leq N}  e ^{\frac{i}{2\sqrt{p}} k_{i,\mu} 
  \Theta^{\mu\nu} k_{j,\nu}\epsilon(u_i-u_j)} \int DX \exp\(-S_{G,
  \Theta=0}[X]+ i\sum_{i=1}^N k_{i,\mu}X^\mu(u_i)\).
$$
We notice that amplitudes involving more general vertex operators (namely waves weighted by a polynomial of derivatives of $X$), would
not allow for contractions coming from the new term in the
propagator. This is the very same situation as in ordinary string
theory, where noncommutativity induces star-products between the
vertex operators, without parturbing the inner structure of any of them.  
From these amplitudes we read off the relevant star-product in
position space, that  actually
 depends on $p$ as well as on the $B$-field:
$$\ast=\exp\( \frac{i}{2\sqrt{p}} \d'_{\mu}
  \Theta^{\mu\nu} \d_{\nu} \)\Big{|}_{x'=x}.$$

\section{Conclusions}
We have established that deformations of the $p$-adic tachyon
Lagrangian by star-products can actually be derived using a
prescription for the coupling of a $p$-adic boundary to a magnetic
field. Moreover, the term in the world-sheet Lagrangian that we
proposed consists of a minimal modification of the action written by
Zhang, that is essentially dictated by the antisymmetry of the
$B$-field.  We observe that the formal limit in which $p$
 goes to 1 yields the usual Moyal product. This provides a sense of
 naturality, that is independent
from the requirement that noncommutativity still appear in
$p$-adic string theory. The way computations are organized is very
similar to the symmetric case, with generalized $p$-adic Gamma-functions
entering the scalar two-point functions, as the close relatives of
$p$-adic Gamma-functions previously used to regularize the propagator.\\

 The non-locality of the effective action, which is manifest even
 without any deformation, made plausible the expectation that
 noncommutativity could still be felt by $p$-adic boundaries. On the
 other hand, the non-locality of the world-sheet action is more
 familiar for the coupling to the $B$-field than it was when the
 symmetric term was exhibited.\\

 The issue of the quadratic extension of ${\rm\bf Q}_p$, that has to be
 chosen so that $\epsilon(-1)=-1$, thus ensuring a non-zero coupling
 to the $B$-field, is a genuinely $p$-adic feature of the
 construction. In usual string theory, the open-string
 world-sheet indeed looks like the complex upper-half plane, the normal
 direction being naturally associated with extension of the field
 ${\bf{R}}$ of the boundary coordinates by square root of unity. Maybe
 the relevance of the extension by $p$ instead of roots of unity is
 related to the fact that the boundary of the $p$-adic world-sheet
 actually consists of leaves of a tree, which come in bouquets of $p$
 leaves, leading to  $p$ links coming from the boundary into the
 interior of the world-sheet, at fixed $p$-adic norm. Making the
 argument more precise would require to obtain our full non-local
 action by integrating out degrees of freedom on the interior of the
 world-sheet. This would compell us to consider $B$ as a genuine
 closed-string field coupling to the whole world-sheet, instead of
 taking the gauge-equivalent picture of a magnetic field coupled to
 the boundary.\\

{\bf Note added.} After this letter had been completed, the work~\cite{GK} appeared, where the same boundary coupling is proposed. It furthermore addresses the problem of invariance of correlation functions under ${\rm GL}(2,\bf{Q}_p)$. This issue is motivated by the decoupling between ordering and algebraic properties occuring in the $p$-adic number field.  

$$ $$

\noindent{{{ \Large \bf Acknowledgements}}}\\ 

\noindent{I} thank Ruben Minasian, Boris Pioline and Pierre Vanhove for
discussions. This work has been partially supported by EC Excellence Grant MEXT-CT-2003-509661.

\end{document}